\documentstyle[fleqn,twoside,gc]{article}

\heads
{D.Yu. Tsipenyuk\foom 1}
{Field Transformation in the Extended Space Model: 
Prediction  and  Experimental Test}

\begin{document}
\twocolumn[
\Arthead{7}{2001}{4 (28)}{336}{338}

\Title{\uppercase{ Field Transformation in the Extended Space Model: \yy
	Prediction  and  Experimental Test }}

\Author{D.Yu. Tsipenyuk\foom 1}
	{General Physics Institute of the Russian Academy of Sciences,
	 38 Vavilova St., Moscow 119991, Russia}

\Rec{07.2001}
\Recfin{11.2001}

\Abstract

] 
\email 1 {tsip@kapella.gpi.ru}

A run of preliminary experiments was carried out to check the prediction of
possible gravitational field generation process arising by stopping of
charged massive particles in a substance  predicted by the recently developed
Extended Space Model (ESM) [1, 3].

ESM is a model of the extended (1+4)-dimensional space $G(T;\vec X,S)$ with the
interval $S$ as the fifth coordinate. Certainly, these five coordinates
satisfy the relation $(ct)^2-x^2-y^2-z^2-s^2=0$. In addition to the Lorentz
transformations $(T;\vec X)$ in (1+3)-dimensional Minkowski space, in
ESM there exist two other transformations in the planes $(T;S)$ and $(\vec
X;S)$. They convert massive particles into massless ones and vice versa.
We also considered the energy-momentum-mass (1+4)D space $G'(E;\vec
P,M)=(E/c;\vec p,mc)$ which is conjugated to the time-coordinates-interval
$(t;x,y,z,s)$ (1+4)D space, and thus a mass $m$ in 5D space $G'(E;\vec P,M)$
corresponds to the interval $s$ in 5D space $G(T;\vec X,S)$. The coordinates
in (1+4)D space $G'(E;\vec P,M)$ satisfy the relation
$E^2-c^2p_x^2-c^2p_y^2-c^2p_z^2-m^2c^4=0$.

Note that this model differs from the analogous 5D theory developed in [6],
in which mass is considered as the fifth coordinate in 5D
time-coordinate-mass (matter) space. In such an approach it is impossible to
build energy-momen\-tum tensor due to mixing of mass coordinate with time and
spatial coordinates.

In the proposed ESM the well-known energy-momentum 4-vector $\widetilde
P(1+3)=(E/c;p_x,p_y,p_z)=(E/c;\vec p)$ in Minkowski space $M(T;\vec X)$
is transformed to the 5-vector $\bar P(1+4)=(E/c;\vec p,mc)$ in the extended
space $G(T;\vec X,S)$ and becomes null for 5-vectors of a massive
particle at rest $(mc;\vec 0,mc)$ as well as for a massless particle
$(\hbar\omega/c; \hbar\omega\vec k/c, 0)$ (here $\vec k$ is the unit vector
in the particle propagation direction).

The basic predictions of general relativity can also be obtained in
ESM. For instance, the following gravitational effects have been considered
in [5]:  the planet escape velocity, starlight redshift and deflection, and
retardation of radar echo from Mars. It has been shown that it is possible
to obtain the same formulae as in the general theory for the magnitudes of
these effects in quite another way in the framework of ESM.

According to the ESM conception, any
external influence applied to any material object can be described as a
change in the appropriate refractive index $n$ at the location of
this object (the introduction of a refractive index for the
gravitational field is well known, see, e.g., [7]).
A change in the refractive index causes reduction of energy, momentum or
mass of the object. Formally, in ESM such a process is described
by rotations in the planes $(T;\vec X),\,(\vec X,S)$ and  $(T;S)$.  We
suppose that the weak gravitational field of a central massive body
generates the refractive index $n(r)=1+\gamma M/rc^2$ in space (where
$\gamma$ is the gravitational constant). The refractive index $n(r)$
depends on the gravitational field strength. This refractive index
determines the motion of both massless (photons) and massive particles in
space. Taking into account the appearance of a refractive index in space
around a massive body, we then apply the technique of rotations in 5D
extended space to calculate various interactions of external bodies with the
gravitational field.

Let us, for instance, briefly show the way of obtaining the second
space velocity in ESM. A massive body at rest is described in our model by
the 5D vector $mc(1;\vec 0,1)$. The motion of a massive body in the
gravitational field along the $x$-axes can be considered in the extended
space $G(T;\vec X,S)$  as a motion in the plane $(\vec X,S)$. Consider the
motion of this massive body from a point where the gravitational field is
absent (where the refractive index is $n=1$) to a point with the
refractive index $n(r)$. This motion is described in EMS by the Euclidean
rotation of a 5D vector in the $(\vec X,S)$ plane:
\bearr
	mc(1;\vec 0,1) \quad \to \quad mc(1;\ -\sin\psi,\ \cos\psi)
\nnn \cm
	=mc\biggl(1;\ -{\sqrt{n(r)^2-1}\over n(r)},\ {1\over n(r)}\biggr).
\earn
Here we take into account that for massless particles (photons) the same
rotation in the plane $(\vec X,S)$ has the form
\[
	{\hbar\omega\over c}(1;\vec 1,0)\to {\hbar\omega\over c}
			(1;\ \vec k\cdot\cos\psi,\ \sin\psi).
\]
The velocity of this particle $v=c\cdot\cos\psi$, hence the refractive index
equals $n=c/v=1/\cos\psi$ [1].

Thus the total velocity gained by a massive particle is
\[
	v=pc^2/E=c\sqrt{n(r)^2-1}/n(r)\simeq\sqrt{2\gamma M/r}.
\]
Here we took into account that in the case of a weak gravitational field
$n(r)\simeq 1$. This value coincides with the escape velocity if we
substitute the radius of the Earth as $r$.

In the framework of ESM, the 5-current $\bar\rho=(\rho,\vec j.j_s)$ is
built instead of the 4-current $\widetilde\rho=(\rho,\vec j)$ as well as
the 5-vector potential $(\varphi,\vec A,A_S)$ instead of $(\varphi,\vec A)$.
With this 5-vector potential it is possible to build field strength tensor
$||F_{ik}||$ whose components are calculated in the usual way  as
$F_{ik}=\d A_i/\d x_k - \d A_k/\d x_i$, $i,k = 0,1,2,3,4$. In the new
$5\times 5$ tensor we have not only the usual electromagnetic components
$E_x=F_{10},\ E_y=F_{20},\ E_z=F_{30},\ H_x=F_{32},\ H_y=F_{13},\
H_z=F_{21}$, but also the new components $Q=F_{40},\ G_x=F_{41},\
G_y=F_{42}$, and $G_z=F_{43}$.  We relate the vector $\vec G$ to the
gravitational field.

In addition, the $5\times5$ second-rank energy-momentum-mass tensor
$||T^{ik}||$  is built in ESM [8] in the same way as  the $4\times4$
energy-momentum tensor is built in the usual $(1+3)$D field theory. The
$||T^{ik}||$ components can be found from the $||F_{ik}||$ components
by well-known formula
\[
   T^{ik}={1\over4\pi}\left(-F^{il}F_l^k+{1\over4}g^{ik}F_{lm}F^{lm}\right),
\]
$i,k,l,m=0,1,2,3,4$.

Besides, an extended set of the Maxwell equations has been obtained [1,2].
This set connects the field strength with the 5-current that gives birth to
the field.

In empty Minkowski (1+3) space $M(T;\vec X)\ (S=0)$, the fields are
independent.  But when we deal with a material medium, which means that the
parameter $S$ becomes nonzero, the two electromagnetic field and a new fields 
$\vec G$ and $Q$ form a unified field, and
their components can interact with each other.

If we rotate  5D vectors, which correspond to 5D-particles in ESM,
in the plane $(T;\vec X)$,  this rotation leads to mixing the particle
momentum and the mass. On the other hand, the rotation in the plane $(T;S)$ 
mixes mass with energy, while the rotation in the plane $(\vec X,S)$ mixes 
momentum with mass.

Further on, we can apply such rotations to the 5D vector potential
$(\varphi,\vec A,A_S)$, the field tensor $||F_{ik}||$, and the fields $\vec
E$, $\vec H$, $\vec G$, $Q$. The fields $\vec E$, $\vec H$, $\vec G$, $Q$
can formally be converted into each other under the same transformations in
the planes $(T;\vec X)$, $(T;S)$ and $(\vec X,S)$ [1, 2].  The
transformation rules are described by introducing the parameters $\vec
v$, $v_s$, $\vec u$. These parameters determine the transition from one
frame of reference to another [1, 2, 9]:

1) rotation in the $(T;\vec X)$ plane is characterised by a velocity
   $\vec v$, and the transformations of fields are
\bearr
	\vec E'=\vec E+(1/c)[v,\vec H],\cm \vec G'=\vec G-(1/c)\vec vQ,
\nnn
	\vec H'=\vec H-(1/c)[v,\vec E],\cm Q'=Q-(1/c)(v,\vec G);
\earn

2) rotations in the plane $(T;S)$ are characterised by the velocity $v_s$
   along the coordinate $S$, and the field transformations are
\bearr
	\vec E'=\vec E+(v_s/c)\vec G,\cm   \vec G'=\vec G+(v_s/c)\vec E,
\nnn
	\vec H'=\vec H,\cm   Q'=Q;
\earn

3) rotations in the plane $(S,\vec X)$ are characterised by the parameter
   $\vec u$, and this  vectorial parameter  describes reduction of the
   refractive index $n$ as a result of motion in the direction of $\vec u$;
   the fields  are transformed according to the relations
\bearr
	\vec E'=\vec E-\vec uQ,\cm    \vec G'=\vec G+[\vec u,\vec H],
\nnn
	\vec H'=\vec H+[\vec u,\vec G], \cm   Q'=Q+(\vec u,\vec E).
\earn

So, we see that the electromagnetic and gravitational fields can be
converted to each other under the corresponding transformation in the planes
$(T;\vec X)$, $(\vec X,S)$, $(T;S)$.

In particular,  when a moving massive charged particle is decelerated,
falling into an external field or substance, it can produce a gravitational
field. Such transformation of fields could, in principle, take place
during various nuclear processes such as gamma or neutron deexcitation,
pair annihilation  or formation and so on.

To check this qualitative prediction, the following experimental setup has
been proposed and recently realized. A relativistic 30\,MeV electron beam
of average power 450\,W from a microtron is stopped in condensed material
(e.g., a tungsten plate). A torsion pendulum with two massive parts
4 kg each and rather a long transverse rod (120 cm) is used to measure the
emerging gravitational field. One of the pendulum shoulders is set very close
to the stopping target --- a possible source of the gravitational field
created by the electrons' deceleration, and another shoulder is apart from
it. As usual, reflection of a laser beam (He-Ne laser with 632.8\,nm
wavelength) from a mirror mounted on a shoulder is used to detect the
possible pendulum deflection. The light spot from the mirror is observed on
a screen located 5\,m apart from the mirror and is registered by a
videorecording system with 15-fold optical gain to increase the measurement
accuracy. The accelerator setup gives us a possibility, very useful for such
a type of experiment, of stopping the electron beam as well in another target,
located near the opposite part of the pendulum shoulder, which could drive
the pendulum rotation in another direction. A more detailed description of
the experiments can be found in Refs.\,[2, 4].

The first tentative experimental results show that there is a correlation
between switchin on the electron beam and the mean deviation of the
pendulum from its equilibrium position as compared with a control run
before and after the switch. It  was also found that the deviation direction
varies depending on which of the pendulum loads was near the stopping
target. An additional evaluation was made by various methods of
statistics, to prove that the deviations detected are statistically
authentic. Thus, by the Pearson criterion, the probability of confident
statistical distinction between the results of control measurements of
pendulum oscillations as compared with the phases of exposure of a stopping
target to relativistic electrons exceeds 99 \%.  The force that could cause
a displacement of the equilibrium position of the pendulum  was also
evaluated. The deflecting force was estimated to be between $10^{-8}$ N and
$10^{-6}$ N.

We carefully checked experimentally various possible sources which could
cause the pendulum deviations: mechanical vibrations due to operation
of water and air pumps, magnetic and electric dc and ac fields,
electrization of different parts of the installation. It was found that all
these sources could not cause the pendulum rotation.

Certainly, these first experimental results on checking the ESM predictions
are of preliminary nature and require much more substantial tests, which
will be a subject of future experiments.

\Acknow{The author expresses his deep gratitude to B.S. Zakirov for help
in realization of the accelerator experiments.}

\end{document}